# Kelvin-Helmholtz vortices in the Martian +E hemisphere downstream of crustal magnetic anomalies

**Junxi Niu[1], Tong Dang[1*], Jiuhou Lei[1*], Binzheng Zhang[2], Wenya Li[3], Binbin Tang[3], Junjie Chen[2], Sudong Xiao[4], Tielong Zhang[4,5]**

[1]Deep Space Exploration Laboratory/School of Earth and Space Sciences, University of Science and Technology of China, Hefei, China.
[2]Department of Earth and Planetary Sciences, The University of Hong Kong, Hong Kong SAR, China.
[3]State Key Laboratory of Solar Activity and Space Weather, National Space Science Center, Chinese Academy of Sciences, Beijing 100190 China.
[4]Harbin Institute of Technology, Shenzhen, China.
[5]Space Research Institute, Austrian Academy of Sciences, Graz, Austria.

## Abstract

At unmagnetized planetary bodies, the Kelvin-Helmholtz (KH) instability at the solar-wind boundary is a key driver of atmospheric escape, fundamentally shaping long-term climate evolution. This shear-driven plasma mixing is theoretically predicted to be asymmetric, strongly favoring the hemisphere where the solar-wind electric field points toward the planet (-E hemisphere). Here we report the observation of a coherent train of fully developed KH vortices in the traditionally unfavorable +E hemisphere of Mars, using in-situ MAVEN measurements. Contemporaneous upstream constraints from Tianwen-1 confirm a steady interplanetary magnetic field, indicating sustained shear-driven growth rather than a transient boundary response. We find that these vortices developed within an expanded, weak-field boundary layer downstream of strong southern crustal magnetic anomalies. Our analysis suggests that regional boundary structuring can override the large-scale hemispheric preference to facilitate localized KH growth. These findings demonstrate that shear-driven escape pathways can operate in traditionally unfavorable regions, altering our understanding of atmospheric volatile loss at weakly magnetized planets.

## Introduction

Understanding the escape of planetary atmospheres is fundamental to reconstruct the history of habitability across the solar system. At planetary bodies lacking a global intrinsic magnetic field, direct interaction with the supersonic solar wind drives substantial atmospheric loss. Mars provides a definitive example: its evolution from a potentially warm and wet environment to today's cold and arid state was profoundly shaped by the cumulative loss of volatiles to space[1,2]. This interaction creates an induced magnetosphere, where the boundary acts as a primary interface for energy and momentum transfer[3–6]. Among the various escape mechanisms, the Kelvin-Helmholtz (KH) instability plays an important role[7–14]. Driven by the velocity shear across the magnetospheric boundary, KH vortices facilitate turbulent plasma mixing, stripping ionospheric material into detached clouds and significantly contributing to global loss rates[15–21]. As a universal shear-driven process across the Solar System, understanding KH behavior at Mars provides a critical observational baseline for evaluating how efficiently the solar wind strips atmospheres from other weakly magnetized planetary bodies.

Corresponding to: Tong Dang (dangt@ustc.edu.cn) and Jiuhou Lei (leijh@ustc.edu.cn)

The prevalence of KH instability at Mars is governed by a well-established hemispheric asymmetry. Earlier studies have suggested that KH waves are preferentially excited in the −E hemisphere, where the solar wind motional electric field points planetward[18,22]. Conversely, the +E hemisphere is dominated by the pickup ion plume[23]. Statistical analyses confirm that the +E boundary is typically thicker with weaker velocity shear, whereas the −E boundary is narrower, sharper, and exhibits stronger shear, conditions highly favorable for KH growth[24]. Kinetic simulations further support this asymmetry, showing that magnetic field pile-up enhances proton drift toward the −E hemisphere, amplifying the velocity shear[18]. Consequently, the +E hemisphere has generally been considered unfavorable for the formation of fully developed KH vortices.

Although single-spacecraft observations have suggested that KH structures may extend beyond traditionally favored regions[16], confirming fully developed roll-up in the +E hemisphere has remained controversial. Recent coordinated MAVEN−Tianwen-1 observations identified 62 KH-related plasma cloud events, yet all were strictly confined to the −E hemisphere, reinforcing the strong statistical hemispheric preference[22]. Crucially, earlier classifications of possible +E events were inferred from non-simultaneous MAVEN solar-wind measurements[16,25–27]. Without contemporaneous upstream constraints, transient variability can complicate the interpretation of boundary structures. Thus, whether non-linear KH roll-up can actually occur in the +E hemisphere under sustained, independently verified solar-wind driving conditions has remained an unresolved observational gap.

Addressing this gap requires independent upstream interplanetary magnetic field (IMF) monitoring coupled with high-resolution in-situ boundary diagnostics. Here, we report a coherent train of five fully developed KH vortices deep within the Martian +E hemisphere using MAVEN[28] observations, directly constrained by contemporaneous IMF measurements from Tianwen-1[29]. Our analysis reveals that this vortex sequence grew under sustained local velocity shear during a highly stable upstream magnetic configuration. We find that this event occurred within a weak-field, expanded boundary layer downstream of strong southern crustal magnetic anomalies. This structural expansion provided a favorable localized environment for KH growth, suggesting that regional crustal field structuring can locally override global electric-field constraints. These findings reveal how regional magnetic anomalies may unlock shear-driven escape in globally unfavorable regions, providing crucial observational benchmarks for global magnetospheric models and our understanding of volatile evolution at weakly magnetized bodies.

## Results

### Event overview

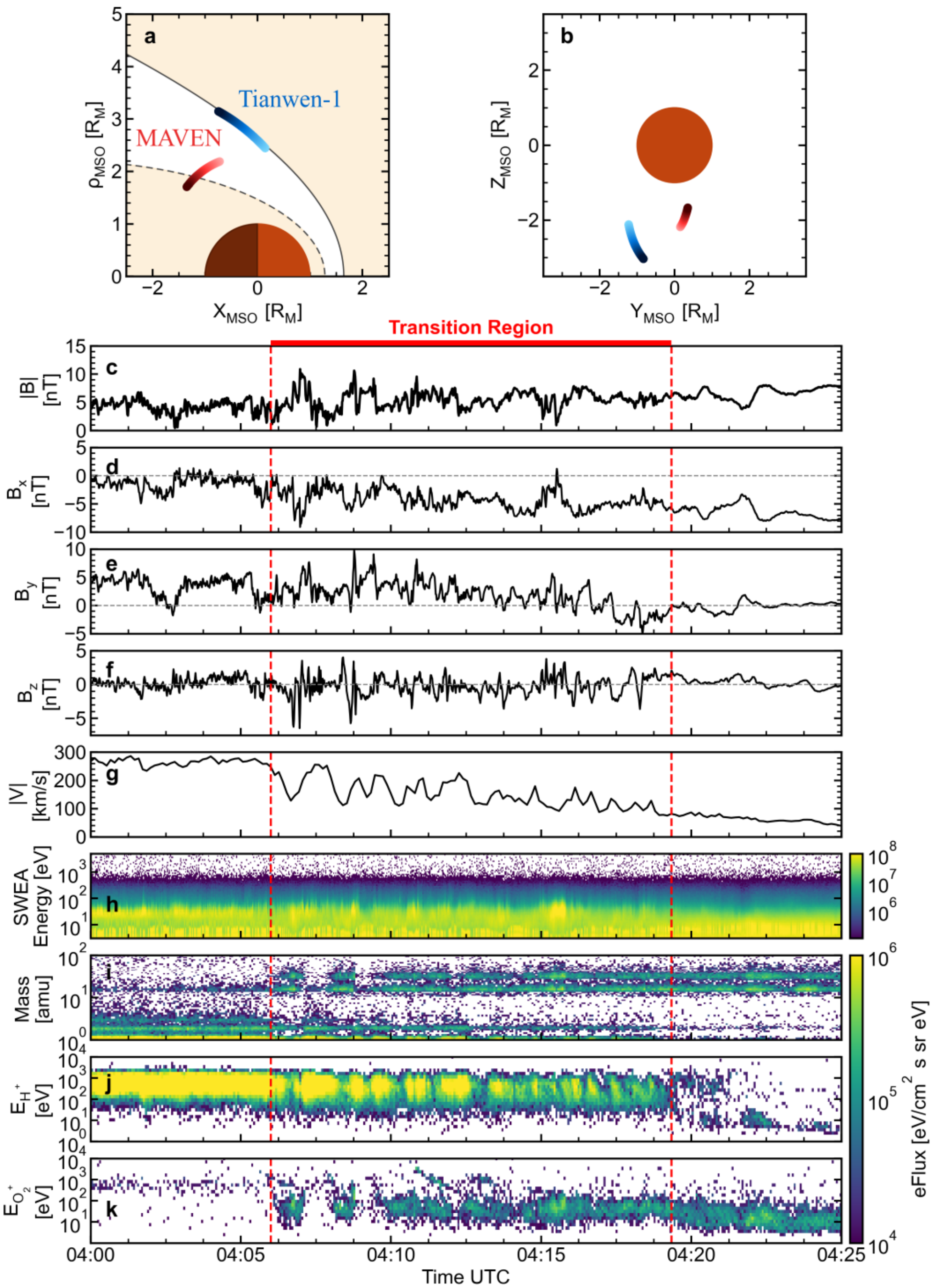

**Figure 1. Overview of the case on May 15, 2023.** Trajectories of MAVEN and Tianwen-1 in Mars Solar Orbital (MSO) coordinates, projected onto the cylindrical $X_{MSO} - \rho_{MSO}$ plane (where $\rho = \sqrt{Y_{MSO}^2 + Z_{MSO}^2}$) **(a)** and the $Y_{MSO} - Z_{MSO}$ plane **(b)**. The solid and dashed black curves represent the nominal bow shock and magnetic pile-up boundary (MPB)[30]. The red color bar and blue color bar represent the time for the spacecraft location shown in **(a)** and **(b)**, with red for MAVEN and blue for Tianwen-1. **(c)** Magnetic field magnitude from the magnetometer (MAG)[31]. **(d–f)** The three components of the magnetic field in MSO coordinates. **(g)** The bulk speed of solar wind protons measured by Solar Wind Ion Analyzer (SWIA)[32]. **(h)** Spectrum of electrons from Solar Wind Electron Analyzer (SWEA)[33]. **(i)** The ion mass spectrum measured by SupraThermal And Thermal Ion Composition analyzer (STATIC)[34]. **(j)** $H^+$ and (k) $O_2^+$ energy spectra from STATIC. The two red vertical dashed lines mark the boundaries of the transition region. Note that the KH interval lies outside the nominal MPB, consistent with a locally elevated boundary.

On May 15 2023, we conducted a coordinated multi-point observation of the Martian plasma environment, using the specific orbital configuration of the Tianwen-1 and MAVEN spacecraft (Fig. 1a and 1b). Tianwen-1 was positioned in the upstream solar wind from 03:22:00 to 04:00:00 UT (see the section "Upstream condition by Tianwen-1"). In contrast, MAVEN remained confined within the bow shock throughout this observation window, precluding direct access to the unshocked solar wind. During the subsequent interval from 04:00:00 to 04:25:00 UT, MAVEN traversed an inbound trajectory in the nightside southern hemisphere, traversing from the magnetosheath across the induced magnetosphere boundary (IMB) also synonymously referred to as the magnetic pile-up boundary (MPB)[35], where it directly encountered the boundary layer instability.

Figure 1c-k presents an overview of this crossing. Prior to 04:06:00 UT, MAVEN was immersed in the magnetosheath, as evidenced by the highly turbulent magnetic field (Fig. 1c-f), an absence of planetary heavy ions (Fig. 1k), and the presence of protons with energies ranging from approximately 50 eV to 2 keV (Fig. 1j), which is consistent with the characteristics of shocked solar wind. By 04:19:20 UT at the end of the interval, MAVEN had entered the induced magnetosphere region, which can be easily identified by a magnetic field with minimal fluctuations (Fig. 1c-f) and the presence of cold (<10 eV) protons (Fig. 1j), the large energy flux of $O_2^+$ (Fig. 1k) and electrons of planetary origin (Fig. 1h).

In the transition region from the magnetosheath to Mars' induced magnetosphere between 04:06:00 and 04:19:20 UT, MAVEN observed multiple quasi-periodic sawtooth-like signatures in the magnetic field (Fig 1c-f), accompanied by quasi-periodic perturbations of $H^+$ velocity (Fig 1g). Further evidence can be found by examining the energy spectrograms of electrons, $H^+$ and $O_2^+$, as well as the ion mass spectrogram (Fig 1h-k). The perturbations alternated between planetary-rich intervals, marked by cold protons and enhanced heavy ions, and magnetosheath-dominated intervals. This pattern indicates the repeated interleaving and mixing of magnetosheath and planetary plasma populations within the boundary structures. Together, these quasi-periodic magnetic and plasma signatures reveal a coherent boundary wave train, while the compositional, velocity, magnetic-field, and pressure diagnostics presented below support its identification as a train of fully developed KH vortices.

## Analysis of the KH vortices

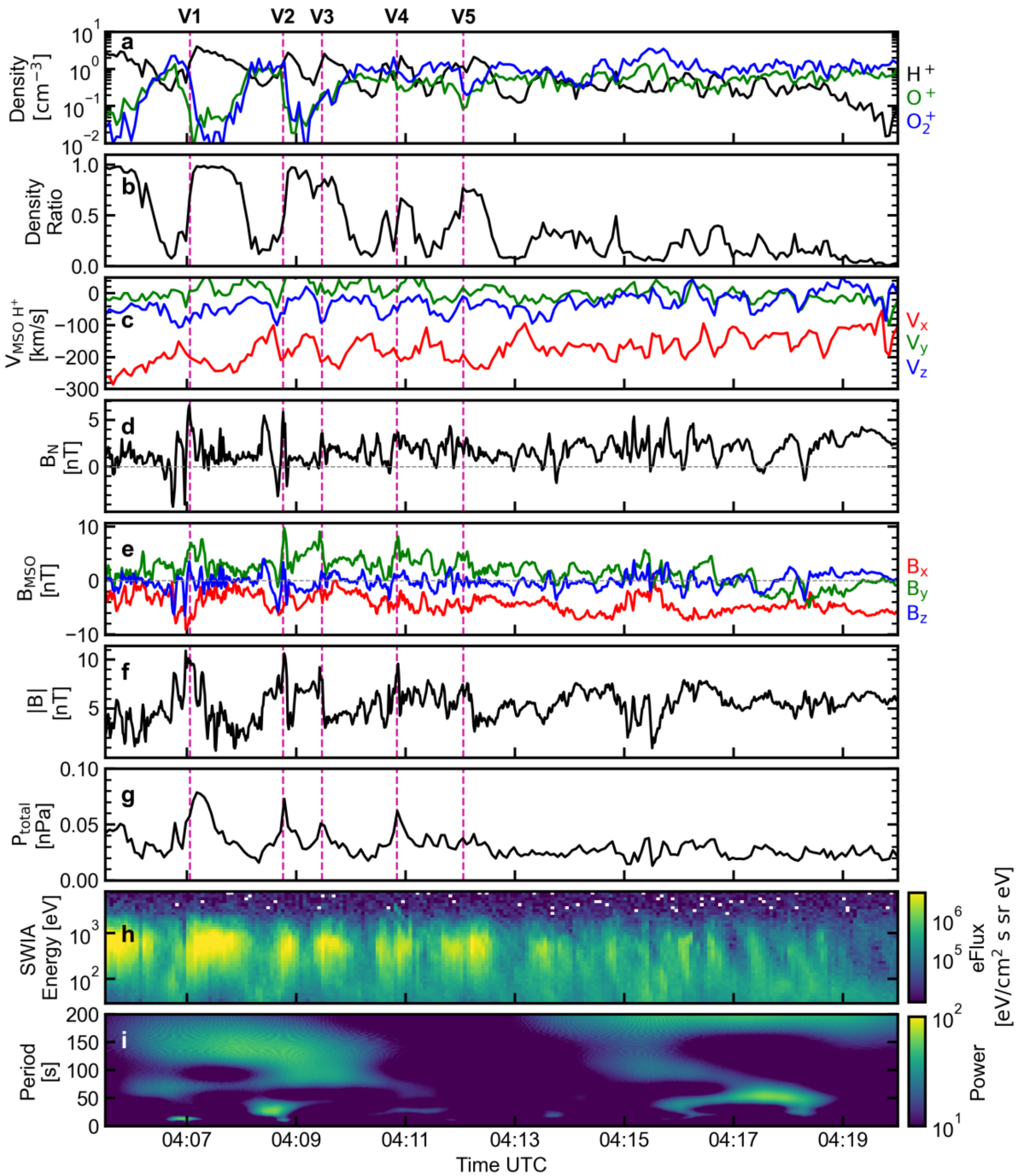

**Figure 2. Plots showing a zoom-in of the transition region for KH vortices. (a)** The number density of $H^+$, $O^+$, and $O_2^+$. **(b)** The ion density ratio $\frac{n_{H^+}}{n_{H^+}+n_{O^+}+n_{O_2^+}}$. **(c)** The three components of $H^+$ velocity in MSO coordinates measured by STATIC. **(d)** The normal component of the magnetic field in LMN coordinates. **(e)** The three components of the magnetic field in MSO coordinates. **(f)** Magnetic field magnitude. **(g)** The total pressure (sum of the magnetic pressure and total plasma pressure). **(h)** The ions energy spectrum measured by SWIA. **(i)** The wavelet power spectrogram of $B_z$. Vertical dashed lines in panels (a–g) indicate the five

KH vortex spine crossings (V1–V5) discussed in the text.

To examine the physical mechanisms behind these wave-like features in greater detail, we perform a zoom-in analysis of the transition region (Fig. 2). The quasi-periodic oscillations observed at the boundary layer provide compelling evidence of the KH instability evolving into a non-linear, fully developed stage. This transition from large-scale surface undulations to localized plasma entrainment is fundamental to understanding the cross-boundary exchange of mass and momentum between the Martian magnetosheath and the ionosphere.

During the crossing, the number densities of magnetosheath-originated $H^+$ and planetary-originated heavy ions ( $O^+$ and $O_2^+$ ) exhibit rhythmic alternations (Fig 2a). This compositional mixing is further quantified by the ion density ratio, $n_{H^+}/\left(n_{H^+} + n_{O^+} + n_{O_2^+}\right)$ (refer to Fig. 2b), which oscillates between values near 1.0 (indicating pure magnetosheath plasma) and significantly lower values (representing an increased planetary component). Such periodic interleaving of distinct plasma populations suggests that MAVEN is traversing the rolled-up interior of the vortices, where mechanical mixing is most intense. These compositional jumps are highly synchronized with perturbations in the $H^+$ velocity components (Fig. 2c), consistent with the velocity shear driving the instability.

The magnetic field morphology provides diagnostic criteria for the non-linear evolution of these vortices. The magnetic field components in MSO coordinates (Fig. 2e) exhibit a distinctive sawtooth-like, bipolar pattern, particularly evident in $B_x$ and $B_y$. These components undergo anti-correlated fluctuations in magnitude and polarity within each wave cycle, a hallmark of the spacecraft encountering the spines of vortices[10,36]. While the sawtooth-like magnetic signatures of KH vortices can superficially resemble the bipolar signatures of magnetic flux ropes, they are physically distinct[37]. Magnetic flux ropes are uniquely characterized by a significant enhancement in magnetic field magnitude $|B|$ along the flux rope axis, coinciding with the center of the bipolar signature[38]. In contrast, the structures observed here lack such axial field enhancement. Furthermore, minimum variance analysis of the magnetic field (MVAB) was performed on the magnetic field across each vortex to characterize the local boundary orientation[39]. For the second bipolar signature corresponding to 04:08:40–04:08:50 UT, the eigenvalue ratios indicate a well-defined normal ($\lambda_2/\lambda_3 = 14.89$). Similar analyses using full-resolution (32 Hz) magnetic field data yield $\lambda_2/\lambda_3$ between 3.09 and 14.89 for the five vortices. Together with the repeated bipolar $B_N$ (Fig. 2d) signatures[12], the absence of a central $|B|$ enhancement, the pressure depletions, and the compositional interleaving, the MVA results support crossings of rolled-up KH structures rather than flux ropes.

Additionally, the total pressure $P_{total}$ (the sum of magnetic and thermal pressures, Fig. 2g) reveals local maxima at the vortex edges and a relative depletion toward the centers. This central pressure deficit is caused by the centrifugal force within the rotating vortex structure, serving as a key mechanical indicator to distinguish KH vortices from magnetic flux ropes, which typically show central pressure enhancement[12,40,41]. The observed pressure perturbation is also consistent with the roll-up criterion, with $\Delta P \approx 0.06\, nPa$ comparable to $\frac{1}{2}\rho(\Delta v)^2 \approx 0.05\, nPa$ , estimated using $n_{H+} \approx 1\, cm^{-3}$ and $\Delta V \approx 259 km/s$.

After 04:13:00 UT corresponding to the latter portion of the transition region, the characteristic peaks in total pressure and magnetic field magnitude largely vanish, accompanied by a reduction in the fluctuation amplitudes of the magnetic field, ion density, and velocity. We suggest that this transition was likely caused by a trajectory effect, where the spacecraft probed the lower flank of the vortices closer to the ionosphere, or alternatively, represents the late-stage dissipation of the vortex train. Furthermore, we identify distinct $H^+$ energy dispersion structures during this interval (see Fig. 2h), notably absent in the heavy planetary ion populations[42]. This energy-time dispersion may reflect kinetic energization processes operating preferentially on the lighter proton population within the turbulent vortex boundary. The detailed acceleration mechanism is beyond the scope of this study and warrants future investigation with higher-cadence particle and field measurements.

**Upstream condition by Tianwen-1**

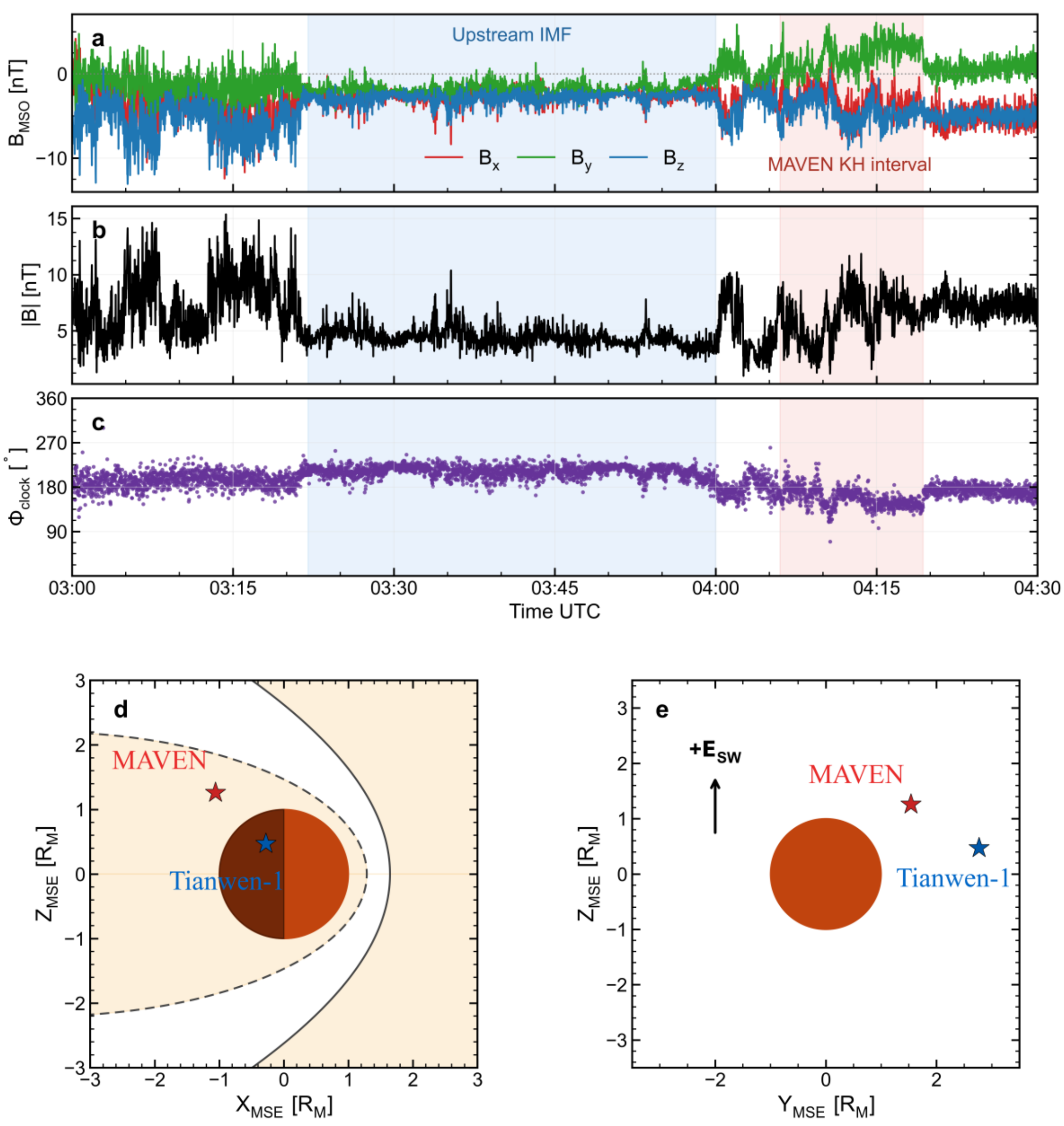

**Figure 3. Tianwen-1's upstream magnetic field measurements indicate the interplanetary context for the MAVEN KH event in the +E hemisphere. (a)** The magnetic fields in MSO coordinates measured by the Mars Orbiter Magnetometer (MOMAG)[43–45] onboard Tianwen-1. **(b)** Magnetic field magnitude. **(c)** The clock angle of the magnetic field. **(d)** Illustrates the projection of MAVEN and Tianwen-1's averaged locations on the $X_{MSE} - Z_{MSE}$ plane and **(e)** the $Y_{MSE} - Z_{MSE}$ plane. The two grey vertical dashed lines mark the boundaries of the upstream IMF. The average MSO positions of Tianwen-1 during the upstream interval and MAVEN during the KH interval were approximately $(-0.28, -1.06, -2.6)\ R_M$ and $(-1.06,\ 0.26, -1.97)\ R_M$. Note that Tianwen-1 projects onto the planetary disk in (d) because of its large out-of-plane position $(Y_{MSE} \approx 2.8\ R_M)$.

To establish the solar wind condition for the observed KH vortices, we utilize near contemporaneous upstream measurements from the Tianwen-1 orbiter (Fig. 3). Before MAVEN encountered the vortex train, Tianwen-1 was positioned in the solar wind, serving as a reliable monitor of the IMF driving the Martian induced magnetosphere. We identify the period from 03:22:00 to 04:00:00 UT as the valid interval for the upstream IMF. On either side of this window, the magnetic field characteristics are consistent with a typical magnetosheath environment, marked by prominent magnetosheath compression and significant magnetic field turbulence (Fig. 3a, b). Within the identified IMF interval, the magnetic field remained stable at approximately $(-2.75, -1.77, -2.85)\ nT$ in MSO coordinates.

Tianwen-1 directly constrains the upstream IMF orientation and temporal stability, but it does not provide upstream plasma moments. To estimate the basic upstream plasma context, we also examined the machine learning-derived estimates of solar wind parameters with associated uncertainties[46]. Around 04:00 UT on 15 May 2023, the model gives a solar wind speed of $\sim 407\ km/s$, proton density of $\sim 7.1\ cm^{-3}$. These values correspond to a dynamic pressure of $\sim 2.0\ nPa$, and an Alfvén Mach number of order 12. We use these proxy derived quantities only to characterize the order of magnitude of the upstream plasma background, whereas the IMF orientation and stability are determined directly from Tianwen-1 measurements. Using this proxy solar wind speed, we estimated the propagation delay between the Tianwen-1 upstream location and the MAVEN KH region. During the upstream interval, Tianwen-1 was located approximately $0.78$ Martian radius sunward of the MAVEN KH region along the $X_{MSO}$ direction. For $V_{sw} \sim 407\ km/s$, this separation corresponds to a propagation time of only $\sim 7\ s$. This delay estimate assumes radial propagation from Tianwen-1 to the MAVEN boundary crossing region. The actual propagation may deviate from this idealized geometry because of IMF orientation, non-radial solar wind flow, or magnetosheath transmission effects. This delay maps the upstream interval onto the boundary nearly instantaneously but leaves six unsampled minutes before the KH onset. The pre-KH magnetosheath field is consistent with continuity of the directly measured IMF orientation during the six-minute gap[25], although it does not provide an independent upstream measurement during the vortex interval.

In general, the KH instability at Mars is theoretically expected to dominate in the −E hemisphere, where the solar wind convective electric field points toward the planet, potentially facilitating the growth of boundary layer instabilities[18,24]. To verify the hemispheric context of this event, we calculated the Mars Solar Electric coordinates (MSE, see "Methods" subsection "Coordinate Systems"). The clock angle, $\varphi$, is defined as the angle between the projected magnetic field and $+Z_{MSO}$ in the $Y_{MSO} - Z_{MSO}$ plane, increasing rotationally toward $+Y_{MSO}$. As shown in Figure 3c, the clock angle of the IMF remained nearly constant at $211.9°$. Under the condition where the upstream IMF has a clock angle of $211.9°$ and assuming that the solar wind velocity is purely along the tailward direction, the MSE unit vectors are determined as: $X_{MSE} = (1, 0, 0)$, $Y_{MSE} = (0, -0.53, -0.85)$, $Z_{MSE} = (0, 0.85, -0.53)$. Figure 3e illustrates the projection of the locations of MAVEN and Tianwen-1 in the $Y_{MSE} - Z_{MSE}$ plane. Using the Tianwen-1 IMF to define the MSE coordinates, we find that the MAVEN KH interval was located well within the $+Z_{MSE}$ (+E hemisphere) (Fig. 3d, e). Although previous observations and simulations indicate that KH growth is more favorable in the −E hemisphere, the present event provides a directly constrained case showing that fully developed vortices can grow deep in the +E hemisphere. Using the MAVEN positions at V1–V5, the $Z_{MSE} = 0$ clock-angle thresholds are $172.7° - 174.1°$. The directly measured pre−event IMF had a mean clock angle of $211.9°$ with a circular standard deviation of $11.5°$. The +E classification is therefore robust to the observed upstream variability and realistic transverse-flow deviations. Although the six-minute gap precludes excluding an unobserved IMF rotation, the pre-KH magnetosheath orientation is consistent with continuity of the same large-scale field sector[25]. It suggests that under specific upstream or localized conditions, the +E hemisphere can also support the evolution of non-linear KH vortices, indicating a more complex interaction between the solar wind and the Martian ionosphere than previously assumed.

**KH conditions and growth rate**

To characterize the observed KH activity, we estimate its period, phase velocity, wavelength, and linear growth rate. As shown in Figure 2i, the KH wave period is near $T_{KH} \approx 100\text{s}$ (range ~80–120 s). Previous magnetohydrodynamic (MHD) simulations have indicated that KH waves propagate at a velocity between the mean velocity $\frac{V_I + V_S}{2}$ and center-of-mass velocity $\frac{\rho_I V_I + \rho_S V_S}{\rho_I + \rho_S}$, where the subscript "I" refers to the ionosphere and "S" indicates magnetosheath[40]. Since the plasma beta during this interval is of order unity, we use this MHD approximation to obtain an estimate of the KH phase speed[15]. Thus, here we assume that $V_{KH}$ is equal to the average of the mean velocity and the center of mass velocity. We use the one-minute periods before and after the transition region to represent the magnetosheath and ionosphere plasmas. Using the plasma moments measured by the STATIC, we calculated the mean mass density (including $H^+$, $O^+$, and $O_2^+$) $\rho_I$ and $\rho_S$ to be $\sim 56\, m_p\, cm^{-3}$ and $\sim 2.6\, m_p\, cm^{-3}$, respectively, where $m_p$ is the mass of a proton. The corresponding mean flow speeds are $V_I$ $\sim 10\, km/s$ and $V_S$ $\sim 269\, km/s$. The mean magnetic field in MSO coordinates are $B_I = (-6.01, -0.21, 0.84)$ $nT$ ($|B_I| \sim 6.08$ $nT$) and $B_S = (-2.18, 2.52, 0.08)$ $nT$ ($|B_S| \sim 3.33$ $nT$). This gives $V_{KH}$ is about $81\, km/s$ (16.7%). The KH wavelength is then estimated as $\lambda_{KH} = V_{KH} \times T_{KH}$, $\lambda_{KH}$ is $\sim 8{,}065\, km$ (19.5%), or $\sim 2.4\, R_M$. This wavelength is of the same order as previously reported Martian KH wavelengths. Such a scale is consistent with the relatively thick transition layer in this event, and with an expanded MPB geometry possibly associated with southern crustal magnetic fields. Uncertainties of the boundary parameters were estimated from the standard deviations within the 1 min averaging windows on each side of the transition region. These uncertainties were propagated to $V_{KH}$, $\lambda_{KH}$, and $\gamma$ using a Monte Carlo calculation in which the input densities, velocities, magnetic fields, and KH period were perturbed within their observed variances.

In the incompressible tangential-discontinuity approximation, the linear growth rate for KH instability is given by the equation[47]:

$$\gamma^2 = \frac{\rho_I \rho_S}{(\rho_I + \rho_S)^2}[\boldsymbol{k} \cdot (\boldsymbol{V_I} - \boldsymbol{V_S})]^2 - \frac{1}{\mu_0(\rho_I + \rho_S)}[(\boldsymbol{k} \cdot \boldsymbol{B_I})^2 + (\boldsymbol{k} \cdot \boldsymbol{B_S})^2] \quad (1)$$

The plasma densities, velocities, and magnetic fields are constrained by the observations, while the magnitude of $\boldsymbol{k}$ is set by the estimated KH wavelength and its direction remains to be determined. By using methodology of Ma[48], we calculated the most KH unstable direction, growth rate $\gamma$, and solid angle $\Omega_{KH}$ over which the instability condition is satisfied. We find that the dominant KH unstable direction is mainly tailward, $(-0.96, -0.077, -0.26)$, consistent with the direction of the strongest flow shear. The corresponding linear growth rate is $\gamma \approx 4.07 \times 10^{-2}\ s^{-1}$ (16.0%), which is larger than previous estimates $1.58 \times 10^{-2}\ s^{-1}$ by Wang et al. and $5.84 \times 10^{-3}\ s^{-1}$ by Poh et al.[19,21]. We also find that the solid angle over which KH instability can occur covers $\sim 86\%$ of $4\pi$ steradians.

The relatively large growth rate and broad unstable solid angle can be understood from the balance between the shear driving and magnetic tension terms in Eq. (1). Because the magnetic stabilization term depends on $(\boldsymbol{k} \cdot \boldsymbol{B})^2$, both the magnetic-field magnitude and its orientation relative to the perturbation wave vector determine the stabilizing effect. For comparable field orientations relative to $\boldsymbol{k}$, reducing the field magnitude from about $20\ nT$ reported in previous KH observations to about $5\ nT$ in this event would reduce the magnetic-tension contribution by a factor of roughly $(5/20)^2 \approx 1/16$. The broad unstable solid angle therefore reflects a locally favorable weak field configuration, not a universal instability over all propagation directions. If such weak field conditions are characteristic of southern crustal field regions, KH growth may be less tightly controlled by the global IMF clock angle than implied by simple hemispheric asymmetry models.

**Potential Atmospheric Oxygen Transport and Escape**

The detachment of fully developed KH vortices and their subsequent evolution into isolated plasma clouds provide an important pathway for the erosion of the Martian ionosphere[17,22]. To assess the potential contribution of the observed +E hemisphere KH wave train to atmospheric loss, we estimate the instantaneous oxygen loss rate ($R_O$) by adopting a simplified cylindrical model consistent with previous studies[9].

In this model, we assume that the ionospheric ions are carried away within a cylindrical structure with a diameter equivalent to the KH wavelength ($\lambda_{KH}$) and a length of approximately $1\ R_M$. The oxygen loss rate is calculated using the following formula[19]:

$$R_O = \pi \left(\frac{\lambda_{KH}}{2}\right)^2 R_M N_O / T_{KH} \quad (2)$$

where $N_O \sim 5 cm^{-3}$ represents the total number density of planetary oxygen ions (including $O^+$ and ${O_2}^+$) and $T_{KH}$ is the observed KH period. Assuming vortex detachment and subsequent escape of the entrained planetary ions, the model yields an event-scale upper-limit oxygen ion loss rate of $\sim 8.65 \times 10^{24}\ s^{-1}$, comparable to previous empirical estimates[3,21,49].

This result suggests that KH-driven plasma detachment in the +E hemisphere could make a non-negligible contribution to atmospheric escape at Mars, although vortex detachment itself is not directly observed here. Quantifying the global role of KH vortices in atmospheric escape will require a broader statistical basis and a better understanding of how individual vortex events evolve into escaping plasma clouds. Furthermore, once detached by KH vortices, these plasma clouds may be further accelerated tailward through a "snowplow" process, in which compressed magnetic fields and electrons mediate momentum transfer from shocked solar wind protons to weakly magnetized planetary heavy ions, thereby enhancing the efficiency of plasma stripping[6,49].

**Role of southern crustal magnetic fields**

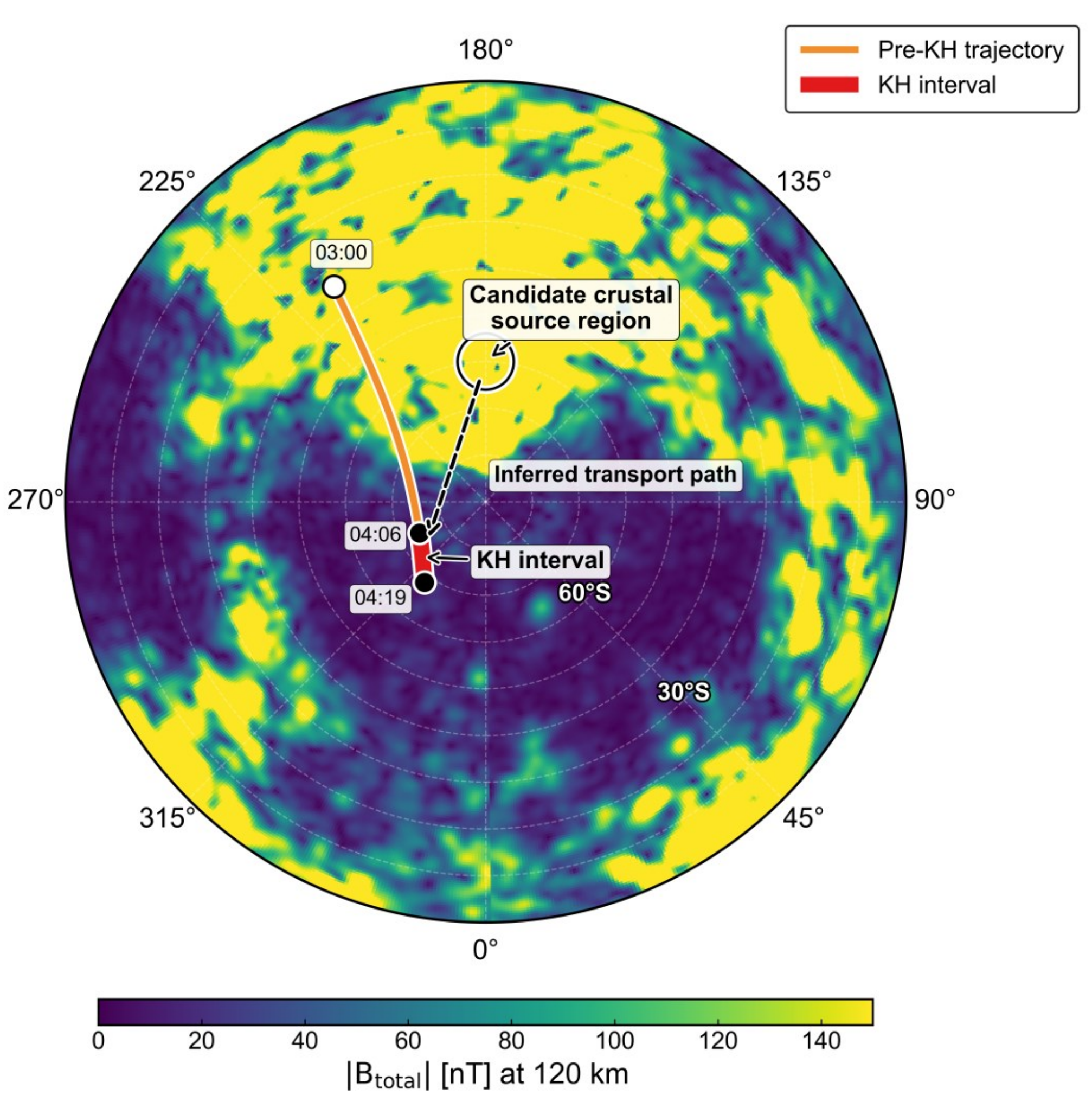

**Figure 4. Spatial relationship between the MAVEN trajectory, the KH interval, and the southern crustal magnetic anomalies.** South polar view of the Martian crustal magnetic field magnitude at 120 km

altitude from the G110 crustal field model[50]. The MAVEN trajectory on 15 May 2023 is divided into the inbound segment before the KH interval and the KH interval. The KH wave train observed during 04:06:00–04:19:20 UT is marked by the red segment and star. The dashed circle indicates a possible crustal source region of boundary perturbations, and the dashed arrow denotes a possible convection path toward the KH observation location. This geometry is consistent with the possibility that the KH vortices developed downstream of boundary structuring associated with strong southern crustal magnetic fields.

Figure 4 shows the south polar view of the Martian magnetic topology together with the MAVEN trajectory on May 15 2023. Before encountering the fully developed vortex train, MAVEN followed an inbound trajectory in the southern hemisphere (Fig. 4). The MAVEN KH crossing was located near a possible convective path from strong southern crustal magnetic anomalies toward the nightside plasma boundary, placing the event downstream of the crustal field region[51,52]. This region hosts the strongest crustal magnetic anomalies on Mars, where interactions between crustal fields and the draped IMF may facilitate magnetic reconnection and contribute to bursty ion escape[53,54]. This geometry suggests that crustal magnetic fields may have influenced the local boundary structure prior to the observation. Previous studies have shown that crustal anomalies modify the Martian plasma boundaries, in particular, the southern (crustal) MPB sits systematically higher and is displaced outward[55,56]. In the present case, such boundary structuring may have provided finite amplitude perturbations that were subsequently amplified by the persistent velocity shear at the MPB. Under this interpretation, crustal fields are best viewed as a possible facilitating factor, rather than as the sole trigger of the observed vortex train.

To examine whether this interpretation is physically reasonable, we estimated the convective timescale required for an initial boundary perturbation to evolve into the observed fully developed vortices. Based on the location of the upstream crustal magnetic anomalies (Fig. 4) and the phase velocity $V_{KH} \sim 81\, km/s$, we estimate a convective timescale of $t_{conv} \approx 70\, s$ for the plasma to transport from the perturbation source to the MAVEN observation point, corresponding to a flow-path separation of approximately 5600 km, $\sim 1.7\, R_M$, between the possible crustal source region and the MAVEN observation point, as indicated by the dashed arrow in Fig. 4. For $\lambda_{KH} \approx 8.07 \times 10^3 km$, $A_c \approx 1 \times 10^3 km$, $\gamma \approx 4 \times 10^{-2} s^{-1}$, and $t_{conv} \approx 70\, s$, the inferred seed amplitude is of order $1 \times 10^2 - 6 \times 10^2 km$ for $t_{nl} \sim 1 - 2\, \gamma^{-1}$ (see “Methods” subsection “Estimation of Vortex Growth from Crustal Perturbations”). This scale is comparable to the order of boundary displacements associated with Martian crustal magnetic anomalies reported in previous studies, suggesting that crustal-field-related boundary roughness could provide a plausible seed for KH growth without requiring unrealistically large initial perturbations[57–60].

We note that the identified KH wave train was situated outside the nominal MPB (Fig. 1a), consistent with an outward displacement of the boundary (away from Mars) in a region influenced by the strong southern crustal anomalies. In addition, the magnetic field magnitudes measured by MAVEN on both sides of the boundary were weak compared with previous Martian KH events[19,21]. Furthermore, the lifting of the MPB by underlying crustal magnetic anomalies may have produced an expanded boundary layer with reduced magnetic flux density, as suggested by the simulation[61]. A related physical tendency has been reported in multifluid simulations of Venus, where ionopause elevation can place the boundary at higher altitudes with reduced collisions and viscosities, thereby enhancing velocity shear and promoting KH growth[62]. Although the physical origin of boundary elevation differs between Venus and Mars, this comparative result supports the broader idea that an elevated or expanded induced plasma boundary can provide more favorable conditions for KH development. As discussed above, crustal-field-related MPB lifting may therefore have favored KH growth by expanding the local boundary layer and reducing the stabilizing effects that would otherwise suppress boundary perturbations.

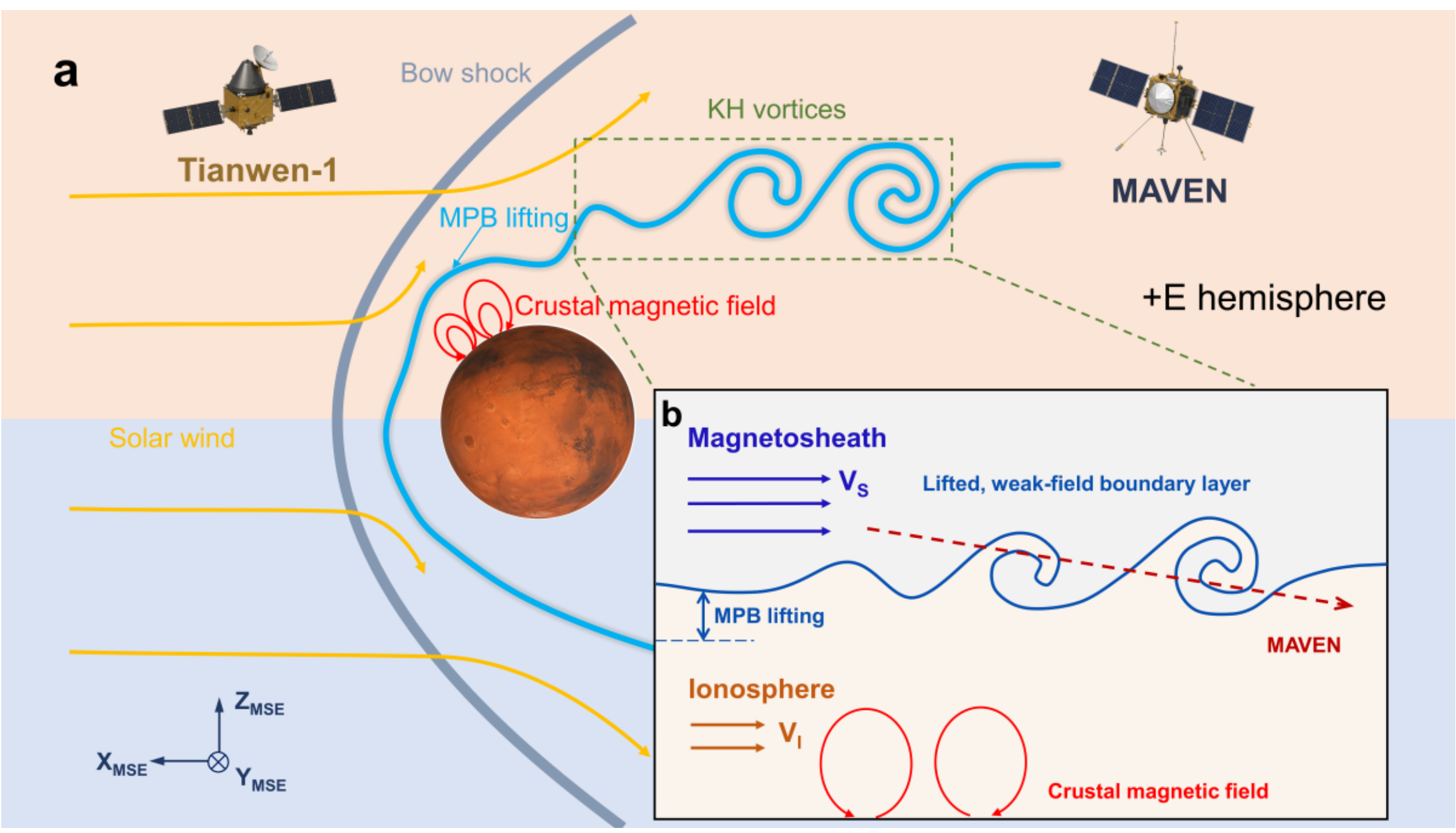

**Figure 5. Schematic interpretation of a crustal-field-related pathway to KH roll-up in the Martian +E hemisphere. (a)** Large-scale geometry of the coordinated MAVEN–Tianwen-1 observations. Tianwen-1 provided upstream interplanetary magnetic field constraints, whereas MAVEN encountered a train of Kelvin–Helmholtz vortices near the induced magnetosphere boundary in the Martian +E hemisphere. The KH interval was located downstream of the southern crustal magnetic field region, where local boundary lifting and expansion may occur. **(b)** Proposed local mechanism. Crustal-field-related boundary structuring may seed finite-amplitude perturbations and produce an expanded transition layer with weak magnetic fields. Under persistent velocity shear between magnetosheath and ionospheric plasma, the reduced magnetic stabilization allows these perturbations to grow into rolled-up KH vortices. This process promotes mixing of magnetosheath protons and planetary heavy ions and provides a pathway for shear-driven plasma transport outside the traditionally favored −E hemisphere.

These observational and quantitative constraints motivate the physical interpretation summarized in Fig. 5. In this scenario, southern crustal magnetic anomalies may have locally lifted and expanded the MPB, generating finite-amplitude boundary perturbations and a weak field transition layer. Persistent velocity shear between the magnetosheath and ionospheric plasma could then amplify these perturbations into rolled-up KH vortices, enabling boundary mixing and plasma transport in the +E hemisphere.

The thickness of the velocity shear layer ($2a$, often denoted as $\Delta$) is the fundamental spatial scale that determines the spectral characteristics and growth of the KH instability. According to the nonlocal linear stability analysis by Miura and Pritchett[63], the KH growth rate depends on the dimensionless parameter $k\Delta$. Linear theory for a hyperbolic tangent velocity profile predicts that KH modes are only unstable within a specific range of $k\Delta <$

2. The fastest growing modes occur for $k\Delta \sim 0.5 - 1.0$. Using the estimated boundary thickness, $\Delta \sim 1152\ km$ (see "Methods" subsection "Estimation of the Transition Layer Thickness"), we obtain $k\Delta \sim 0.9$, within the unstable range and near the fastest-growing regime. Because the crossing samples an already-developed wave train, $\Delta$ is an effective upper bound and $k\Delta$ serves as a consistency check rather than a precise measurement. This consistency is nevertheless important for understanding why fully developed vortices could occur in the +E hemisphere. Previous statistics indicate that +E hemisphere boundaries are typically thicker and have weaker shear, both of which tend to suppress KH growth[24]. Here, however, the combination of $k\Delta \sim 0.9$ and a velocity contrast of $\Delta V \sim 259\ km/s$ is consistent with locally favorable conditions for KH growth despite the global hemispheric preference. This local configuration is consistent with MPB lifting and boundary expansion associated with southern crustal magnetic anomalies.

## Discussion

The central result of this study is that non-linear KH vortices were observed in the traditionally less favorable +E hemisphere, constrained by steady upstream IMF measurements from Tianwen-1. This finding is particularly notable in light of recent coordinated MAVEN−Tianwen-1 observations that identified 62 KH-related plasma cloud events, all confined to the −E hemisphere[22]. The absence of this event from the survey of Zhang et al. may reflect their upstream-selection criteria[22]. Although our direct upstream IMF constraint ends six minutes before the KH interval, Tianwen-1 measurements from the preceding orbit also showed a stable clock angle near $260°$, corresponding to the same +E configuration. Together, these observations support the robustness of the +E classification. The present event therefore provides a directly constrained exception to the strong statistical −E preference and shows that electric-field hemisphere is not an absolute control on KH development. Instead, local boundary conditions, including boundary displacement, shear-layer structure, and crustal-field-related boundary perturbations, may substantially modify the stability of the induced magnetosphere boundary.

The upstream constraint is important for distinguishing this event from transient-driven boundary disturbances. The Martian plasma environment can respond strongly to upstream variability, including interplanetary coronal mass ejections, dynamic pressure enhancements, IMF rotations, and transient foreshock-related structures[4,5,61]. The persistent +E vortex train reported here differs from the transient −E escape enhancement of Zhang et al.[5], for which Tianwen-1 served as a spatial reference rather than a direct upstream monitor. Such transients can generate localized ion escape or boundary perturbations that may superficially resemble shear-driven structures if upstream conditions are not measured simultaneously[25–27]. In our event, the steady IMF recorded by Tianwen-1 argues against a large-scale IMF rotation as the primary trigger. In addition, the coherent, quasi-periodic train of five vortices observed by MAVEN is difficult to reconcile with an isolated foreshock transient or a single upstream pressure pulse. These constraints support a persistent, shear-driven origin for the observed structures, consistent with the classical KH interpretation of rolled-up vortices at planetary plasma boundaries[10–12]. Nevertheless, because Tianwen-1 did not measure upstream plasma moments, we cannot completely exclude smaller-scale solar wind dynamic pressure variations. The key point is that the available upstream magnetic field observations and the regular vortex morphology make a purely transient interpretation unlikely.

This event should not be taken to imply that the previously proposed hemispheric asymmetry is unimportant. Previous observations and simulations indicate that KH growth at Mars is generally more favorable in the −E hemisphere, where the solar-wind motional electric field, boundary structure, and proton velocity shear tend to promote instability[18,22,24]. Our observations instead suggest that this hemispheric preference represents a background tendency rather than an absolute rule. In the present case, the KH wave train occurred downstream of strong southern crustal magnetic anomalies, in a weak-field and expanded boundary layer. Such crustal-field-related boundary structuring may have provided finite-amplitude perturbations and reduced magnetic stabilization, allowing persistent velocity shear to amplify the perturbations into rolled-up vortices. In this sense, crustal magnetic fields are best interpreted as a possible facilitating factor, not as the sole trigger of the event. This local interpretation is consistent with recent statistics showing that crustal fields do not control the global occurrence of Martian plasma clouds[22]. It also differs from the stabilizing role proposed for other southern boundary configurations[21].

The observed event also has implications for atmospheric escape from Mars. The estimated event-scale oxygen loss rate indicates that fully developed KH vortices in the +E hemisphere can produce localized planetary ion transport comparable to previously reported Martian plasma cloud escape events[6,21,49]. Therefore, shear-driven plasma escape may not be restricted to the traditionally favored −E hemisphere. Instead, regions affected by crustal-field-related boundary expansion may provide additional channels for plasma mixing and detachment. However, this single event does not justify a quantitative reassessment of the global escape rate. The global importance of KH-driven escape depends on the occurrence frequency, spatial distribution, mass loading, detachment efficiency, and subsequent tailward acceleration of plasma clouds.

It should also be pointed out that several limitations need to be considered when interpreting this event. Because the analysis is based on a single well-constrained case, it cannot determine how commonly +E hemisphere KH vortices occur under similar crustal field and upstream conditions. Although the upstream IMF was directly measured by Tianwen-1, the upstream density and velocity were inferred from proxy estimates rather than measured in-situ, leaving some uncertainty in the solar wind plasma context[46]. In addition, the proposed role of southern crustal magnetic fields is supported by spatial geometry, boundary displacement, weak magnetic fields, and growth-timescale estimates, but establishing a direct causal link will require targeted numerical simulations. Future statistical studies combining MAVEN and Tianwen-1 observations[28,29], together with hybrid or multifluid simulations that include realistic crustal magnetic fields[55,57,60], can test how often regional boundary structuring overcomes the hemispheric preference. The present event nevertheless demonstrates that the electric-field hemisphere is not an absolute control on KH development, and that regional magnetic structure can shape shear-driven transport and long-term volatile loss at weakly magnetized planetary bodies[1,2].

## Methods

### Instruments

This study uses coordinated in-situ observations from the MAVEN and Tianwen-1 orbiters to characterize the plasma boundary event and its upstream context. Magnetic field measurements at MAVEN are taken from the Magnetometer (MAG) instrument[31], which consists of dual tri-axial fluxgate sensors. Tianwen-1 magnetic field measurements are provided by the Mars Orbiter Magnetometer (MOMAG)[43–45]. In this study, we use 1 s magnetic field data from both spacecraft, along with 32 Hz MAVEN MAG data. Ion distributions are obtained from MAVEN's Solar Wind Ion Analyzer (SWIA)[32], which measures ion fluxes from 5 eV to 25 keV using a toroidal electrostatic analyzer, and the Suprathermal and Thermal Ion Composition (STATIC) instrument[34], which combines a toroidal "top hat" electrostatic analyzer with a time-of-flight (TOF) section to resolve ion species (e.g., $H^+$, $O^+$, $O_2^+$) over an energy range of 0.1 eV to 30 keV, with both ion instruments providing nominal 4-second time resolution. Solar Wind Electron Analyzer (SWEA) measures a full three-dimensional distribution of electrons with range from 3 eV/q to 4.6 keV/q within 2 s[33].

Although Tianwen-1 carries a plasma instrument, upstream plasma moments suitable for this event are not currently available. Therefore, in this study Tianwen-1 is used primarily to provide upstream magnetic field measurements during the interval relevant to the MAVEN boundary crossing. These magnetic field observations are sufficient to constrain the IMF orientation, assess its short-term stability, and determine the MSE hemispheric context of the event, which are the primary roles of Tianwen-1 in the present analysis.

**Coordinate Systems**

We employ two Cartesian coordinate systems to analyze the spatial geometry and electromagnetic environment. The first is the Mars Solar Orbital (MSO) coordinate system. In this frame, $\boldsymbol{X_{MSO}}$ points from the center of Mars toward the Sun, $\boldsymbol{Y_{MSO}}$ is opposite to the component of Mars' orbital velocity perpendicular to $\boldsymbol{X_{MSO}}$, and $\boldsymbol{Z_{MSO}}$ completes the right-handed system perpendicular to the orbital plane. To better understand the interaction between the solar wind and the planetary environment, particularly regarding the orientation of the induced magnetosphere, we also adopt the Mars Solar Electric (MSE) coordinate system. In the MSE system, the $\boldsymbol{X_{MSE}}$ axis is defined to be antiparallel to the upstream solar wind velocity vector (approximated here as aligned with $\boldsymbol{X_{MSO}}$). The $\boldsymbol{Z_{MSE}}$ axis aligns with the direction of the upstream solar wind convective electric field ($\boldsymbol{E_{SW}} = -\boldsymbol{V_{SW}} \times \boldsymbol{B_{IMF}}$), and $\boldsymbol{Y_{MSE}}$ completes the right-handed system.

Because Tianwen-1 does not measure the upstream plasma velocity, the MSE transformation in this study assumes a purely anti-sunward solar-wind flow, that is, $\boldsymbol{V_{SW}} \parallel -\boldsymbol{X_{MSO}}$. This approximation introduces uncertainty into the exact orientation of $\boldsymbol{Z_{MSE}}$ if the solar wind contains a non-negligible transverse flow component. For a transverse velocity of order $10-15\%$ of the anti-sunward flow speed, the inferred $\boldsymbol{Z_{MSE}}$ direction would rotate by only $\sim 6° - 9°$. Since MAVEN's average position during the KH interval lies well within the +E hemisphere and far from the $Z_{MSE} = 0$ boundary, this uncertainty does not alter the hemispheric classification of the event. The identification of this interval as a +E hemisphere KH event is therefore robust to realistic deviations of the solar wind flow direction from the ideal anti-sunward assumption.

**Estimation of the Transition Layer Thickness**

To quantify the physical scale of the KH instability region, we estimated the thickness (Δ) of the transition layer (magnetic pile-up boundary, MPB) using the spacecraft's trajectory and a geometric model of the boundary. The thickness is determined by the projection of the spacecraft's velocity onto the boundary's normal direction during the crossing interval.

The duration of the boundary crossing, $\Delta t$, is defined as the interval between the transition region ($t_{start}$ to $t_{end}$). The average spacecraft velocity vector ($\mathbf{V_{sc}}$) in the Mars Solar Orbital (MSO) frame is calculated from the displacement between these two points:

$$\mathbf{V_{sc}} = \frac{\boldsymbol{R}(t_{end}) - \boldsymbol{R}(t_{start})}{\Delta t} \tag{3}$$

where $\boldsymbol{R}$ represents the position vector of MAVEN.

The normal vector of the MPB ($\mathbf{n}$) is derived using a conic section model, which describes the boundary surface as a symmetric rotation around the Mars-Sun axis. The surface is defined by the implicit equation:

$$F(x,y,z) = (x - x_F)^2 + y^2 + z^2 - [L - \epsilon(x - x_F)]^2 = 0 \tag{4}$$

where $x_F$ is the focus position, $L$ is the semi-latus rectum, and $\epsilon$ is the eccentricity. Based on the crossing location, we utilized specific parameters for the MPB by Trotignon et al.[30]. The unit normal vector $\mathbf{n}$ at the crossing midpoint is then calculated as the normalized gradient of the surface function:

$$\nabla F = \begin{pmatrix} \frac{\partial F}{\partial x} \\ \frac{\partial F}{\partial y} \\ \frac{\partial F}{\partial z} \end{pmatrix} = \begin{pmatrix} -2(\epsilon^2 - 1)(x - x_F) + 2\epsilon L \\ 2y \\ 2z \end{pmatrix} \tag{5}$$

$$\mathbf{n} = \pm \frac{\nabla F}{|\nabla F|} \tag{6}$$

The effective thickness ($\Delta$) is finally obtained by multiplying the absolute normal velocity of the spacecraft by the crossing duration:

$$\Delta = |\mathbf{V_{sc}} \cdot \mathbf{n}| \cdot \Delta t \tag{7}$$

Under the stationary nominal-boundary assumption, this projection gives an apparent geometric thickness. Because the crossing samples an already-developed vortex train, $\Delta$ is treated as an effective upper bound rather than a direct measurement of the undisturbed shear-layer thickness.

**Estimation of Vortex Growth from Crustal Perturbations**

To evaluate whether upstream crustal magnetic anomalies could plausibly provide the initial perturbation for the observed KH event, we employ a simplified convective growth estimate. We define the convective timescale, $t_{conv}$, as the time required for a plasma cloud to travel from the inferred crustal anomaly source region to the spacecraft observation point,

$$t_{conv} = \frac{\Delta L}{V_{KH}} \tag{8}$$

where $\Delta L$ is the distance along the flow streamline between the upstream crustal anomaly and the observation site, and $V_{KH}$ is the phase velocity of the KH wave. Because the derived growth rate $\gamma$ is obtained from linear theory, it is applicable only while the boundary displacement remains small compared with the wavelength. For a sinusoidal interface perturbation $\eta = Aexp(ikx + \gamma t)$, the small-amplitude assumption requires $kA \ll 1$, or equivalently a small interface slope. We therefore take the onset of non-linear evolution to occur when $kA$ becomes of order unity. This gives a characteristic non-linear threshold amplitude[47,64],

$$A_c \sim k^{-1} = \frac{\lambda_*}{2\pi} \tag{9}$$

where $\lambda_*$ is used only as an order-of-magnitude characteristic scale of the mature KH structure. This estimate does not assume that the observed wavelength is strictly conserved throughout the full non-linear evolution; it simply marks the point at which the linear, small-amplitude approximation is expected to break down.

During the linear stage, the perturbation is assumed to grow exponentially from an initial seed amplitude $A_0$ to $A_c$,

$$A_c = A_0 e^{\gamma t_{lin}} \tag{10}$$

where $t_{lin}$ is the duration of the linear growth stage. The remaining time is assigned to the non-linear stage, such that

$$t_{conv} = t_{lin} + t_{nl} \tag{11}$$

where $t_{nl}$ is the non-linear evolution timescale. Combining Eqs. (10) and (11) gives

$$A_0 = A_c \exp[-\gamma(t_{conv} - t_{nl})] \tag{12}$$

In this framework, the inferred $A_0$ should be regarded as an order-of-magnitude estimate. We then compare the resulting range of seed amplitudes with the characteristic scale of MPB perturbations associated with crustal magnetic anomalies to assess whether a crustal-field-related seeding is physically plausible.

### Data availability

The MAVEN data used to generate the plots are available via the Planetary Data System at https://pds.nasa.gov or the MAVEN Science Data Center at https://lasp.colorado.edu/maven/sdc/public/. The Tianwen-1 MOMAG data sets are publicly available at http://space.ustc.edu.cn/dreams/tw1_momag/.

### Code availability

The Python code used to generate Figs. 1-4, together with the processed input data, environment specification and reproduction instructions, is available at Zenodo at https://doi.org/10.5281/zenodo.21821801.

## Author contributions

J.N. performed the data analysis, prepared the figures and wrote the original draft. T.D. conceived the study, supervised the work and revised the manuscript. J.L. contributed to the scientific interpretation and supervision. B.Z., W.L., B.T., J.C., S.X. and T.Z. contributed to data interpretation, discussion of the results and revision of the manuscript. All authors reviewed and approved the final manuscript.

## Competing interests

The authors declare no competing interests.